\documentclass[12pt]{article}
\usepackage{amsfonts}
\usepackage[dvips]{graphicx}
\usepackage{latexsym,amsmath,amssymb,graphics,stmaryrd}
\usepackage{cite}
\usepackage{array}
\usepackage{subfigure}
\usepackage{multirow}
\usepackage{epsfig}
\usepackage[dvips]{graphicx}
\usepackage[dvips]{color}
\usepackage{pstricks}
\usepackage{pst-node}
\usepackage{pst-plot}
\usepackage{dsfont}
\usepackage{amsthm}
\usepackage{graphics}
\usepackage{subfigure}
\usepackage{fancyhdr}
\usepackage[dvips]{color}

\usepackage{feynmp}

\fancypagestyle{plain}{
\fancyhf{}
\newcommand{\fpage}{\iffloatpage{}{\thepage}}
\fancyfoot[C]{\fpage}

}

\newcommand{\be}{\begin{equation}}
\newcommand{\ee}{\end{equation}}

\newcommand{\col}{~,}
\newcommand{\pnt}{~.}

\newcommand{\la}{\lambda}
\newcommand{\hla}{\hat{\lambda}}

\newcommand{\NN}{\mathcal{N}}
\newcommand{\DD}{\mathcal{D}}
\newcommand{\s}{\sigma}

\newcommand{\pfour}[4]{{}\{#1,#2,#3,#4\}{}}

\newcommand{\ptwo}[2]{{}\{#1,#2\}{}}
\newcommand{\pone}[1]{{}\{#1\}{}}

\newlength{\neglength}
\newlength{\diameter}
    
\DeclareMathOperator{\Tr}{Tr}

\DeclareMathOperator{\perm}{P}

\numberwithin{equation}{section}
\newlength{\unit}
\psset{xunit=\unit,yunit=\unit,runit=\unit}
\newlength{\linew}
\psset{linewidth=\linew}
\begin{document}
%\immediate\write18{open -g -a spires \jobname.tex}
\pagestyle{empty}

\begin{flushright}\footnotesize
\texttt{UUITP-18/10} %\\
%\texttt{NBI-??}
\vspace{0.8cm}\end{flushright}

\renewcommand{\thefootnote}{\fnsymbol{footnote}}
\renewcommand{\theequation}{\arabic{equation}}
\renewcommand{\thefootnote}{\arabic{footnote}}
\setcounter{footnote}{0}
\setcounter{equation}{0}

\begin{center}
{\Large\textbf{\mathversion{bold} A limit on the ABJ model}}

\bigskip
{\it Contribution to the proceedings of QTS6}

\vspace{1.5cm}

\textrm{J.~A.~Minahan$^{1}$, O.~Ohlsson Sax$^{1}$ and
C.~Sieg$^{2}$}
\vspace{8mm}

\textit{$^{1}$ Department of Physics and Astronomy, Uppsala University\\
SE-751 08 Uppsala, Sweden}\\
\texttt{joseph.minahan@fysast.uu.se, olof.ohlsson-sax@physics.uu.se} \vspace{3mm}

\textit{$^{2}$The Niels Bohr International Academy\\ The Niels Bohr Institute\\
Blegdamsvej 17,
 DK-2100, Copenhagen \O, Denmark }\\
\texttt{csieg@nbi.dk}
\vspace{3mm}

%%%%%%%%

\par\vspace{1cm}

\textbf{Abstract} \vspace{5mm}

\begin{minipage}{14cm}

There is a large amount of evidence that the  ABJM model is integrable in the planar limit.  Less clear is whether or not the ABJ model is integrable.  Here we investigate a limit of the ABJ model in the weak coupling limit where one 't Hooft parameter is much greater than the other.  At the 4 loop level in the $SU(2)\times SU(2)$ sector the anomalous dimensions of single trace operators map to two Heisenberg spin chains with nearest neighbor interactions with an overall coefficient that is a function of one of the 't Hooft parameters.   We conjecture the form of this function and show that is consistent with observations about the ABJ model concerning unitarity and parity, including strong coupling statements.  

\end{minipage}

\end{center}

\vspace{0.5cm}

%%%%%%%%%%%%%%%%%%%%%%%%%%%%%%%%%%%%%%%%%%%%%%%%%%%%%%%%%%%%%%%%%%%%%%%%%%%%%%%%

%\newpage
%\setcounter{page}{1}
\renewcommand{\thefootnote}{\arabic{footnote}}
\setcounter{footnote}{0}
\setcounter{equation}{0}

%%%%%%%%%%%%%%%%%%%%%%%%%%%%%%%%%%%%%%%%%%%%%%%%%%%%%%%%%%%%%%%%%%%%%%%%%%%%%%%

The ABJM model  \cite{Aharony:2008ug} is a $U(N)_k\times U(N)_{-k}$ superconformal Chern-Simons theory with $\NN=6$ supersymmetry.  In the planar limit the  two point functions of single trace operators map to a spin chain of alternating type that is believed to be integrable \cite{Minahan:2008hf,Gaiotto:2008cg,Gromov:2008qe,
Bak:2008cp}.  The integrability is remarkably similar to that found for $\NN=4$ super Yang-Mills.  However, one important difference is the presence of an unknown function $h^2(\la)$, where $\la$ is the 't Hooft coupling $\la=N/k$.  This function appears in the Bethe equations as well as the spin chain magnon dispersion relation 
\be\label{E}
E(p)=\sqrt{Q^2+4h^2(\lambda)\sin^2\tfrac{p}{2}}-Q\,,
\ee
where $Q$ is an $R$-charge.  In $\NN=4$ SYM, $Q=1$ and $h^2(\la)=\frac{\la}{4\pi}$.  For the ABJM model $Q=1/2$, but $h^2(\la)$ is only known at its asymptotic limits, where for weak coupling $h^2(\lambda)=\lambda^2+{\rm O}(\lambda^4)$, while for strong coupling $h^2(\lambda)=\tfrac{1}{2}\lambda -\frac{\ln2}{\sqrt{2}\pi} \sqrt{\la}+ {\rm O}(1)$.

The ABJ model \cite{Aharony:2008gk} is an interesting extension of ABJM, where now the gauge group is $U(M)_k\times U(N)_{-k}$ but the amount of supersymmetry is unchanged.  In the planar limit there are now two 't Hooft constants, $\la=M/k$, $\hla=N/k$, but whether or not ABJ is integrable is not clear.  It turns out that at the two-loop level, the lowest order of perturbation theory, the anomalous dimension mixing matrix still maps to an integrable spin-chain \cite{Bak:2008vd,Minahan:2009te}.  However, the corresponding string dual for this theory is type IIA on $AdS_4\times CP^3$ with a $b$-flux through a $CP^1$ on the $CP^3$.  This $b$-flux corresponds to a $\theta$-angle for the world-sheet theory, where $\theta=2\pi(N-M)/k+\pi$  \cite{Aharony:2009fc}, which generically breaks parity.  The $\theta$-angle cannot affect the classical integrability \cite{Arutyunov:2008if,Stefanski:2008ik} since it does not change the equations of motion.  However, at the quantum level it is expected to have an effect.  This is what happens for  the $O(N)$ model, where the two parity preserving values, $\theta=0,\pi$ are known to be integrable.  However, the IR behavior for these two values of $\theta$ are completely different, for example the $\theta=0$ point has a gap while the $\theta=\pi$ point is gapless.  It is not clear at all how to smoothly interpolate between these two IR points by varying $\theta$, leaving the integrability in doubt.

For the purpose of this note let us ask  what  the behavior would be if the planar ABJ model {\it were} integrable  for any value of $\la$ and $\hla$.  In this case it seems that the only possible modification to the ABJM results is to replace the function  $h^2(\la)$ with $h^2(\la,\hla)$ in the dispersion relation and the Bethe equations.  At the two loop level it has been shown that it simply replaces $\la^2$ with $\la\hla$ \cite{Bak:2008vd,Minahan:2009te}.  This two loop behavior motivates us to define two new variables $\bar\la$ and $\s$ where
\begin{equation}\label{barlambdasigmadef}
\bar\lambda=\sqrt{\lambda\hat\lambda}\col\qquad
\sigma=\frac{\lambda-\hat\lambda}{\bar\lambda}\,,
\end{equation}
and then write $h^2(\bar\la,\s)$ in terms of these two variables.  

Since parity transformations in the ABJM/ABJ theories   include the map $A_\mu\leftrightarrow\hat A_\mu$, the ABJ theories are not in general parity invariant.  One might expect this would translate into parity violation in the spin chain, manifested by different dispersion relations for odd and even site magnons, or equivalently with   $h^2(\bar\la,\s)\ne h^2(\bar\la,-\s)$.  However, at least up to and including the four-loop level there turns out to be  no parity violation.

In \cite{Minahan:2009aq,Minahan:2009wg} we reported on a four loop calculation to compute the first corrections to $h^2(\la,\hla)$.  Since $h^2(\la,\hla)$ can be extracted from the magnon dispersion relation, it is not necessary to consider the most general spin chains.  Instead we can restrict ourselves to an $SU(2)\times SU(2)$ subsector which is closed.  The magnons on the odd sites are the excitations in one $SU(2)$ while the magnons on the even sites are the excitations for the other $SU(2)$.  Furthermore, the odd-site magnons do not interact with the even-site magnons until the eight loop level so at  four loops they can be treated as noninteracting.  

The function $h^2(\bar\la,\s)$ is assumed to have the following form consistent with parity conservation,
\begin{equation}\label{h4ansatz}
\begin{aligned}
h^2(\bar\lambda,\sigma)=\bar\lambda^2+\bar\lambda^4h_4(\sigma)\col\qquad
h_4(\sigma)=h_4+\sigma^2h_{4,\sigma}
\col
\end{aligned}
\end{equation}
A $\s^4$ term would also be consistent with parity, but a quick inspection of possible Feynman diagrams shows that such a term does not appear.  

The 4-loop correction $h_4(\s)$ is extracted from 4-loop diagrams by considering the expansion of the dispersion relation
\begin{equation}\label{Eexpansion}
\begin{aligned}
E(p)
&=2\bar\lambda^2(1-\cos p)
+2\bar\lambda^4(h_4(\sigma)-3+(4-h_4(\sigma))\cos p-\cos2p)
+\mathcal{O}(\bar\lambda^6)\,.
\end{aligned}
\end{equation}
These terms arise from  the permutation structures that sit  in the spin chain Hamiltonian, or equivalently the dilatation operator $D$.  From equation (\ref{Eexpansion}) we see that $D$ has the form
\begin{equation}
D=L+\bar\lambda^2D_2+\bar\lambda^4D_4(\sigma)+\mathcal{O}(\bar\lambda^6)
\,,
\end{equation}
with
\begin{equation}\label{dilatation}
\begin{aligned}
D_{2,\text{even}}&=\pone{}-\pone{1}\col\\
D_{2,\text{odd}}&=\pone{}-\pone{2}\col\\
D_{4,\text{odd}}(\sigma)
&=(h_4(\sigma)-4)\pone{}+(6-h_4(\sigma))\pone{1}-\ptwo{1}{3}-\ptwo{3}{1}
\col\\
D_{4,\text{even}}(\sigma)
&=(h_4(\sigma)-4)\pone{}+(6-h_4(\sigma))\pone{2}-\ptwo{2}{4}-\ptwo{4}{2}
\pnt
\end{aligned}
\end{equation}
The notation $\{\dots\}$ describes the permutation structures, where
\begin{equation}\label{permstruc}
\begin{aligned}
\pfour{a_1}{a_2}{\dots}{a_m}=\sum_{i=1}^L\perm_{2i+a_1\,2i+a_1+2}\perm_{2i+a_2\,2i+a_2+2}\dots\perm_{2i+a_m\,2i+a_m+2}\col
\end{aligned}
\end{equation}

Inspection of (\ref{dilatation}) shows that $h_4(\s)$ can be extracted from $D$ by considering those contributions  that involve a single exchange.  There are still dozens of Feynman diagrams to consider that are listed in \cite{Minahan:2009aq,Minahan:2009wg}.  Here we just report the result, \begin{equation}
\begin{aligned}\label{h4result}
h_4=-4\,\zeta(2)\col\qquad 
h_{4,\sigma}=-\zeta(2)\pnt
\end{aligned}
\end{equation}
Note that the result for $h_4$ is consistent with a dampening of the quadratic behavior at weak coupling to the linear behavior at strong coupling.  Also note that both $h_4$ and $h_{4,\s}$ have highest {transcendentality}. 

In \cite{Minahan:2009aq,Minahan:2009wg} it was also shown that one could take  $\la\gg\hla$ and  consider the rescaled dilatation operator
 $\mathcal{D}=\hat\lambda^{-1}(D-2L)$, which at the 4-loop level reduces to
 \begin{equation}\label{localham}
\begin{aligned}
\mathcal{D}=\lambda(1-\zeta(2)\lambda^2)
(2\pone{}-\pone{1}-\pone{2})
\end{aligned}
\end{equation}
for the  $SU(2)\times SU(2)$ sector.  But this  is simply the Hamiltonian for two decoupled Heisenberg spin chains.  We further believe that the form of $\DD$ persists to higher loop level, that is the Hamiltonian will still be two decoupled  Heisenberg chains with an overall coefficient that is some function of $\la$, $f(\la)$.

The form of the 4-loop expansion is very suggestive and we conjecture that the function is\footnote{We thank N. Nekrasov for suggesting this possibility to us.}
\begin{equation}\label{function}
f(\la)=\frac{1}{\pi}\,\sin(\pi\,\la)\,.
\ee
Let us now motivate why $f(\la)$ in (\ref{function}) has the right behavior.  

First, as was argued in \cite{Aharony:2008gk}, unitary ABJ theories with gauge group  $U(N+\ell)_{k}\times U(N)_{-k}$ do not exist if $\ell>k$.  In terms of the 't Hooft parameters this corresponds to $\la>\hla+1$.  In the limit where 
{$\la\gg\hla$} this translates into $\la>1$ violating the unitary bound.   If we now consider the function in (\ref{function}) we see that it changes sign as $\la$ passes through $1$.  But if the overall coefficient is negative, then so are the anomalous dimensions.  Included in these is  the single trace-operator ${\Tr}[(Y^1Y^\dag_3Y^2Y^\dag_4-Y^1Y^\dag_4Y^2Y^\dag_3)]$ which is a superpartner of ${\Tr}[Y^AY^\dag_A]$ that too must have the same  negative anomalous dimension.  But then  ${\Tr}[Y^AY^\dag_A]$ would have a total dimension that is slightly less than 1, violating the unitary bound for superconformal field theories in 3 dimensions \cite{Minwalla:1997ka}.  Hence, the spin-chain also indicates that there is no such unitary theory.

A second test of (\ref{function}) is related to the first and relies on parity.  Based on the brane picture there are reasons to believe that {the} $U(N+\ell)_k\times U(N)_{-k}$ theory is equivalent to $U(N)_k\times U(N+k-\ell)_{-k}$ \cite{Aharony:2008gk}.  This corresponds to the 't Hooft pair equivalence $(\la,\hla)\equiv(\hla,2\hla+1-\la)$.  However, if parity is preserved at least for the spin chain calculations then one should find the same anomalous dimensions for  $(2\hla+1-\la,\hla)$.  In the limit where {$\la\gg\hla$} this indicates an equivalence between the theory with $\la$ and the theory with $1-\la$, which is consistent with the $f(\la)$ in (\ref{function}).

Of course more evidence is needed to confirm the form of (\ref{function}).  A six-loop calculation to compute the next term in the expansion is a daunting proposition but may be feasible, since the number of Feynman graphs that must be computed is quite restricted.  We are presently investigating such a calculation.

%%%%%%%%%%%%%%%%%%%%%%%%%%%%%%%%
\subsection*{Acknowledgments}
We would like to thank N.~Nekrasov for helpful discussions. 
The research of J.\ A.\ M.\ is supported in part by the
Swedish research council and the STINT foundation.  J.\ A.\ M.\ thanks the
CTP at MIT   for kind
hospitality  during the course of this work. 
%%%%%%%%%%%%%%%%%%%%%%%%%%%%%%%%%%

%\clearpage

\footnotesize
\bibliographystyle{JHEP}
\bibliography{references}

\providecommand{\href}[2]{#2}\begingroup\raggedright\begin{thebibliography}{10}

\bibitem{Aharony:2008ug}
O.~Aharony, O.~Bergman, D.~L. Jafferis, and J.~Maldacena, {\it
  {${\mathcal{N}}\!=6$ Superconformal Chern-Simons-Matter Theories, M2-Branes
  and Their Gravity Duals}},  {\em JHEP} {\bf 10} (2008) 091,
  [\href{http://xxx.lanl.gov/abs/0806.1218}{{\tt arXiv:0806.1218}}].

\bibitem{Minahan:2008hf}
J.~A. Minahan and K.~Zarembo, {\it {The Bethe Ansatz for Superconformal
  Chern-Simons}},  {\em JHEP} {\bf 09} (2008) 040,
  [\href{http://xxx.lanl.gov/abs/0806.3951}{{\tt arXiv:0806.3951}}].

\bibitem{Gaiotto:2008cg}
D.~Gaiotto, S.~Giombi, and X.~Yin, {\it {Spin Chains in ${\mathcal{N}}\!=6$
  Superconformal Chern-Simons-Matter Theory}},  {\em JHEP} {\bf 04} (2009) 066,
  [\href{http://xxx.lanl.gov/abs/0806.4589}{{\tt arXiv:0806.4589}}].

\bibitem{Gromov:2008qe}
N.~Gromov and P.~Vieira, {\it {The All Loop AdS4/CFT3 Bethe Ansatz}},  {\em
  JHEP} {\bf 01} (2009) 016, [\href{http://xxx.lanl.gov/abs/0807.0777}{{\tt
  arXiv:0807.0777}}].

\bibitem{Bak:2008cp}
D.~Bak and S.-J. Rey, {\it {Integrable Spin Chain in Superconformal
  Chern-Simons Theory}},  {\em JHEP} {\bf 10} (2008) 053,
  [\href{http://xxx.lanl.gov/abs/0807.2063}{{\tt arXiv:0807.2063}}].

\bibitem{Aharony:2008gk}
O.~Aharony, O.~Bergman, and D.~L. Jafferis, {\it {Fractional M2-Branes}},  {\em
  JHEP} {\bf 11} (2008) 043, [\href{http://xxx.lanl.gov/abs/0807.4924}{{\tt
  arXiv:0807.4924}}].

\bibitem{Bak:2008vd}
D.~Bak, D.~Gang, and S.-J. Rey, {\it {Integrable Spin Chain of Superconformal
  U(M)Xu(N) Chern- Simons Theory}},  {\em JHEP} {\bf 10} (2008) 038,
  [\href{http://xxx.lanl.gov/abs/0808.0170}{{\tt arXiv:0808.0170}}].

\bibitem{Minahan:2009te}
J.~A. Minahan, W.~Schulgin, and K.~Zarembo, {\it {Two Loop Integrability for
  Chern-Simons Theories with ${\mathcal{N}}\!=6$ Supersymmetry}},  {\em JHEP}
  {\bf 03} (2009) 057, [\href{http://xxx.lanl.gov/abs/0901.1142}{{\tt
  arXiv:0901.1142}}].

\bibitem{Aharony:2009fc}
O.~Aharony, A.~Hashimoto, S.~Hirano, and P.~Ouyang, {\it {D-Brane Charges in
  Gravitational Duals of 2+1 Dimensional Gauge Theories and Duality Cascades}},
   \href{http://xxx.lanl.gov/abs/0906.2390}{{\tt arXiv:0906.2390}}.

\bibitem{Arutyunov:2008if}
G.~Arutyunov and S.~Frolov, {\it {Superstrings on $\mathrm{AdS}_4$ $\times$
  Cp$^3$ as a Coset Sigma-Model}},  {\em JHEP} {\bf 09} (2008) 129,
  [\href{http://xxx.lanl.gov/abs/0806.4940}{{\tt arXiv:0806.4940}}].

\bibitem{Stefanski:2008ik}
B.~Stefanski, jr., {\it {Green-Schwarz Action for Type IIA Strings on
  $\mathrm{AdS}_4\times $ Cp$^3$}},  {\em Nucl. Phys.} {\bf B808} (2009)
  80--87, [\href{http://xxx.lanl.gov/abs/0806.4948}{{\tt arXiv:0806.4948}}].

\bibitem{Minahan:2009aq}
J.~A. Minahan, O.~Ohlsson~Sax, and C.~Sieg, {\it {Magnon Dispersion to Four
  Loops in the Abjm and Abj Models}},
  \href{http://xxx.lanl.gov/abs/0908.2463}{{\tt arXiv:0908.2463}}.

\bibitem{Minahan:2009wg}
J.~A. Minahan, O.~Ohlsson~Sax, and C.~Sieg, {\it {Anomalous Dimensions at Four
  Loops in ${\mathcal{N}}\!=6$ Superconformal Chern-Simons Theories}},
  \href{http://xxx.lanl.gov/abs/0912.3460}{{\tt arXiv:0912.3460}}.

\bibitem{Minwalla:1997ka}
S.~Minwalla, {\it {Restrictions Imposed by Superconformal Invariance on Quantum
  Field Theories}},  {\em Adv. Theor. Math. Phys.} {\bf 2} (1998) 781--846,
  [\href{http://xxx.lanl.gov/abs/hep-th/9712074}{{\tt hep-th/9712074}}].

\end{thebibliography}\endgroup

\end{document}

%%% Local Variables: 
%%% mode: latex
%%% TeX-master: "integrals"
%%% End: 